\documentclass[lettersize,journal]{IEEEtran}
\usepackage{amsmath,amsfonts}
\usepackage{algorithmic}
\usepackage{algorithm}
\usepackage{array}
\usepackage[caption=false,font=normalsize,labelfont=sf,textfont=sf]{subfig}
\usepackage{textcomp}
\usepackage{stfloats}
\usepackage{url}
\usepackage{verbatim}
\usepackage{graphicx}
\usepackage{cite}
\usepackage{booktabs}   % 用于 \toprule, \midrule, \bottomrule
\usepackage{amssymb}    % 用于 \checkmark
\usepackage{amsmath}    % 用于 \mathcal
\usepackage{multirow}
\usepackage{graphicx}   % 用于调整表格宽度

\usepackage{siunitx}
\usepackage{amsmath,amssymb,amsfonts}
\usepackage{wrapfig}
\usepackage{threeparttable}
\usepackage{multirow}
\usepackage{makecell}

\usepackage{orcidlink}
\usepackage{hyperref}

\begin{document}

\title{Geometric Correction of Side-Scan Sonar Images with Image-Consistent Attitude Refinement}

\author{Can Lei ~\orcidlink{0009-0006-1301-8615},
        Valerio Franchi ~\orcidlink{0000-0002-9592-9618},
        Hayat Rajani ~\orcidlink{0000-0002-2541-2787},~\IEEEmembership{Member,~IEEE},
        Nuno Gracias ~\orcidlink{0000-0002-4675-9595},
        Rafael Garcia ~\orcidlink{0000-0002-1681-6229},
        and Huigang Wang        ~\orcidlink{0000-0001-6225-7205},~\IEEEmembership{Member,~IEEE}
        % <-this % stops a space
\thanks{Can Lei and Huigang Wang are with the School of Marine Science and Technology, Northwestern Polytechnical University, Xi’an 710072, China. Huigang Wang is also with the Research \& Development Institute of Northwestern Polytechnical University in Shenzhen, Shenzhen 518057, China.}% <-this % stops a space

\thanks{Valerio Franchi, Hayat Rajani, Nuno Gracias and Rafael Garcia are with the Computer Vision and Robotics Research Institute (ViCOROB) of the University of Girona, Spain. This work was conducted while Can Lei was on a research stay at ViCOROB, Spain}% <-this % stops a space

\thanks{This work was partly supported by the Spanish government through projects ASSiST (PID2023-149413OB-I00) and IURBI (CNS2023-144688). This work was also supported by the National Natural Science Foundation of China (62171368) and Science, Technology and Innovation of Shenzhen Municipality (Grant Nos. JCYJ20241202124931042, ZDCYKCX20250901093900002).}% <-this % stops a space
\thanks{Corresponding author: Huigang Wang (e-mail: wanghg74@nwpu.edu.cn).}}

\maketitle

\begin{abstract}

Side-scan sonar (SSS) images are susceptible to motion-induced geometric distortion, which degrades their reliability for seabed interpretation and downstream tasks. Existing correction methods either exploit image-domain consistency without adequately preserving global geometric referencing, or rely on navigation-based geocoding whose effectiveness is limited when recorded attitude and motion fail to capture ping-scale perturbations. To address this issue, we propose a geometric correction method for SSS images with image-consistent attitude refinement. The core idea is to refine the yaw-pitch sequence used in geocoding by explicitly linking stripe-wise distortion patterns in dual-sided waterfall images to geometric deformation modes. Specifically, a navigation-derived macro-scale attitude baseline is fused with image-inferred microscopic perturbations, where port-starboard symmetry is used to separate pitch-related common-mode responses from yaw-related differential-mode responses. The refined attitude is then incorporated into a physically geocoding framework with track-aligned gridding and normalized-convolution-based hole completion to generate the corrected image. Experiments on real SSS datasets from different sonar platforms and environments show that the proposed method reduces inter-ping misalignment, local stretching, and structural discontinuity, and improves local geometric consistency under both degraded-attitude and cross-dataset evaluation settings, demonstrating its effectiveness for geometrically consistent SSS correction.
\end{abstract}


\section{Introduction}
\IEEEPARstart{S}{ide}-scan sonar (SSS) is a widely used acoustic imaging modality for large-area seafloor observation in turbid underwater environments, providing high-resolution seabed imagery for geomorphological interpretation \cite{A1}, target detection \cite{A2} and classification \cite{A3}. However, because SSS is inherently a line-scan imaging system, its raw waterfall image is formed by stacking successive echo profiles along-track \cite{A4}, and is therefore highly sensitive to time-varying platform motion and pose parameters. Attitude oscillation, speed fluctuation, trajectory perturbations, timing mismatch between the sonar and navigation systems, and filtering-induced phase lag can all disrupt the geometric consistency of adjacent pings on the seafloor projection. These effects often appear as inter-ping misalignment, along-track stretching \cite{A5}, and local compression, degrading the reliability of downstream tasks such as terrain matching, and thus necessitating robust geometric correction.

Existing correction strategies generally rely on adjustment in the image-domain \cite{A6} or geocoding \cite{A7} with measured navigation and attitude parameters. The former can exploit local image consistency but lacks absolute spatial constraints, making global drift difficult to avoid. The latter preserves global georeferencing through an explicit imaging geometry model, but its performance depends critically on the fidelity of the input motion and attitude data. In real surveys, navigation measurements are often affected by high-frequency noise, drift, timing delay, and filtering artifacts \cite{A8}, so that they preserve the coarse motion trend but fail to accurately represent the ping-scale perturbations that directly affect line-scan image formation. As a result, residual stripe stretching and local misalignment may still remain after standard geocoding.

The inherent limitations of these two approaches highlight a critical complementarity in SSS geometric correction. While navigation data provides the absolute reference necessary for global spatial consistency, the ping-scale perturbations responsible for residual distortions are better captured by local image consistency than by noisy navigation records. Motivated by this observation, this paper proposes a geometric correction method for SSS images with image-consistent attitude refinement, where the pose input of geocoding is refined rather than replaced. Specifically, a refined yaw-pitch sequence is constructed by combining a navigation-derived macro-scale baseline with image-inferred microscopic perturbations extracted from dual-sided waterfall images under port-starboard constraints. The refined attitude is then incorporated into geocoding, followed by track-aligned gridding and adaptive completion of empty cells, thereby improving the local geometric consistency of corrected SSS imagery while preserving global georeferencing fidelity. The main contributions of this paper are summarized as follows:
\begin{itemize}
    \item We formulate SSS geometric correction as a geocoding-oriented attitude refinement problem, where the goal is not to recover the true physical motion state itself, but to estimate an image-consistent yaw-pitch sequence for geometrically consistent SSS correction.

    \item We propose an image-consistent microscopic attitude inference strategy that extracts stripe-wise distortion evidence from dual-sided waterfall images and exploits the port-starboard response pattern to separate the refinement cues associated with yaw and pitch.

    \item We develop a refined-attitude geocoding framework that combines navigation-preserved global geometric stability with image-driven local correction, and further improves image completeness and structural continuity through track-aligned gridding and adaptive hole filling.

    \item We validate the effectiveness, robustness, and generalization capability of the proposed method through extensive experiments on multiple real-world datasets collected from different sonar systems and environments.
\end{itemize}

\section{Related Work}

Existing studies on SSS geometric correction can be broadly grouped into two categories: image-based correction methods that infer distortion directly from the sonar image itself, and geocoding-based methods that rely on navigation, attitude, and auxiliary measurements to establish geometric consistency in a geographic frame.

\subsection{Image-Based Geometric Correction}

Early attempts at SSS geometric correction considered compensating motion-induced distortion directly in the image domain, without relying on external navigation or attitude measurements. A representative example is the work of Cobra et al. \cite{A80}, who estimated geometric distortion by cross-correlating adjacent scan lines under the assumption that platform motion irregularities manifest as relative misalignment between neighboring pings. By iteratively compensating these inter-line shifts, his method reduced local distortion and identified reverse scanning directions using image information alone.

To the best of our knowledge, however, image-based geometric correction has seen very limited follow-up in the SSS literature and has not become a widely adopted correction framework. A key limitation is that inter-ping misalignment in the image domain does not uniquely correspond to platform motion, since similar appearance changes may also arise from seabed relief \cite{A9}, scattering variation, or texture transition. In addition, purely image-driven line-by-line compensation is prone to error accumulation along track, making it difficult to preserve global geometric consistency or achieve absolute georeferencing. As a result, image-domain information is better viewed as a source of local geometric evidence than as a complete standalone solution for SSS geometric correction.

\subsection{Geocoding-Based Geometric Correction}

Most subsequent studies have adopted geocoding-based correction frameworks, in which slant-range measurements are projected into a physical coordinate system using navigation, attitude, and auxiliary sensor data \cite{A100}. Because such methods provide explicit geometric constraints, they can suppress the global drift that is difficult to avoid in purely image-based correction and have therefore become the dominant paradigm in practical SSS processing.

Early studies mainly focused on correcting slant-range distortion and speed-induced deformation under simplified geometric assumptions. Sheffer et al. \cite{A10} proposed a correction strategy based on the flat-seabed assumption, followed by speed compensation using GPS-derived platform velocity. This method reduced distortions caused by range geometry and speed variation with relatively low computational cost, but its applicability is limited in areas with significant terrain relief. To improve correction under nonflat seabed conditions, Ye et al. \cite{A11} incorporated bathymetric measurements by constructing seabed terrain models for more accurate slant-range correction. This improved geometric fidelity over varying topography, but the use of averaged platform velocity within GPS update intervals makes the method sensitive to abrupt speed changes and maneuvering conditions.

More recent studies have extended geocoding-based correction through georeferenced projection, interpolation, and task-oriented preprocessing. Franchi et al. \cite{A12} projected sonar measurements onto a two-dimensional georeferenced grid and reconstructed the image along smoothed trajectories using gaussian interpolation, which reduced motion-induced distortion and improved downstream classification, although it may increase sensitivity to noise in sparsely sampled regions and computational cost in high-resolution surveys. Cui et al. \cite{A13} addressed visual artifacts in the water-column region after correction through self-similarity-based restoration, but this type of method mainly acts as postprocessing rather than physically grounded correction. For efficiency-oriented applications, Huang et al. \cite{A14} adopted flat-seabed-based slant-range correction as preprocessing for AUV target recognition; however, such simplified strategies generally neglect attitude-induced deformation and complex seabed topography.

In summary, geocoding-based methods remain a practical and widely adopted solution for SSS geometric correction, but existing approaches treat auxiliary navigation and attitude data as absolute ground truth, making them ill-equipped to handle the inevitable dynamic measurement uncertainties discussed earlier. Consequently, local geometric inconsistencies may persist even after geocoding. This motivates the introduction of image-based constraints derived from the acoustic data itself, so that global geometric referencing and local distortion correction can be considered in a more unified manner.

\begin{figure}[htbp]
	\centering
	\includegraphics[width=1\linewidth]{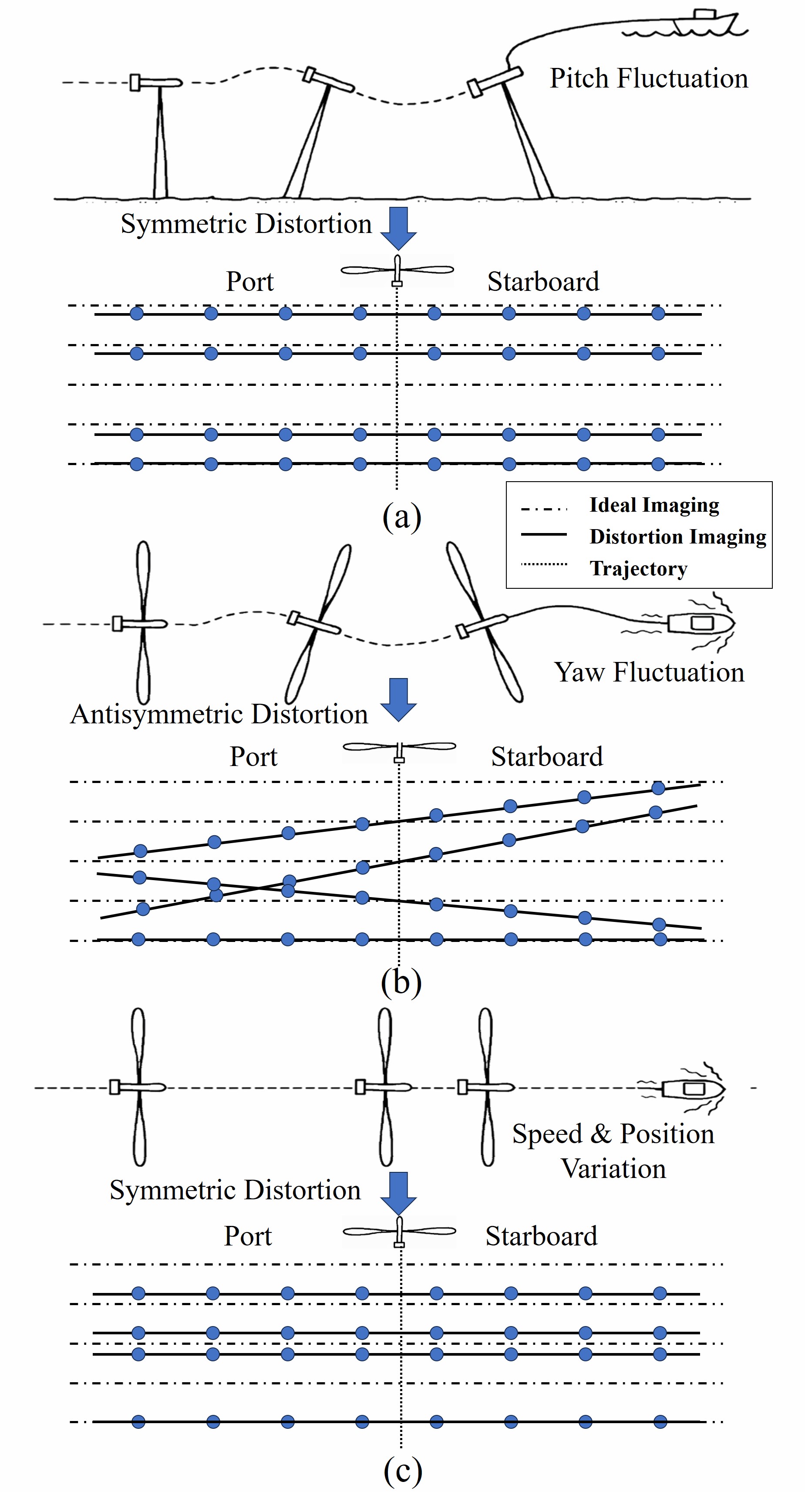}
	\caption{Schematic illustration of the observable distortion patterns in dual-sided SSS waterfall imagery. (a) Pitch fluctuation causes symmetric along-track distortion on the port and starboard sides. (b) Yaw fluctuation causes antisymmetric mirrored distortion, advancing one side while delaying the other. (c) Towfish speed and position variations generate a symmetric stacking-step disturbance that is observationally equivalent to a pitch-related along-track scale perturbation.}
	\label{theory}
\end{figure}	

\section{Method}

\subsection{Modeling Assumptions and Equivalent Parameterization}

During SSS acquisition, the extrinsic state of the $n$-th ping is denoted as

\begin{equation}
\boldsymbol{\eta}(n)=\left(\mathbf{p}_{\mathrm{tow}}(n),\,
\tilde{\boldsymbol{\Theta}}(n),\,V(n),\,h(n)\right),
\end{equation}
where $\mathbf{p}_{\mathrm{tow}}(n)\in\mathbb{R}^{2}$ denotes the towfish position,
$\tilde{\boldsymbol{\Theta}}(n)\in\mathbb{R}^{3}$ denotes the attitude vector (roll/yaw/pitch),
$V(n)\in\mathbb{R}$ denotes the towfish speed, and
$h(n)\in\mathbb{R}$ denotes the altitude above the seafloor.

\subsubsection{From Full Extrinsics to an Equivalent Refinement Model}

To obtain an observable image-driven refinement model, we adopt the following assumptions and reductions:

\begin{itemize}

\item Roll mainly modulates backscatter intensity via incidence-angle variation \cite{A15} and has a minor contribution to the along-track misalignment considered here, hence it is not treated as a primary refinement variable.

\item Altitude $h(n)$ can be robustly obtained by bottom-line tracking \cite{A16}, so we keep it fixed during refinement to avoid ill-conditioned coupling with other extrinsics.

\item As shown in Fig. \ref{theory}(c), fluctuations in $\mathbf{p}_{tow}(n)$ and $V(n)$ produce the same symmetric along-track distortion pattern in waterfall imagery \cite{A17} as a pitch-induced disturbance, and are therefore absorbed into the pitch channel.

\end{itemize}

With these treatments, residual geometric distortions are parameterized by two observable components: the pitch refinement, capturing along-track equivalent scale inconsistency, and the yaw refinement, capturing heading-projection deflection. Notably, we seek image-driven attitude refinement to improve the geometric consistency of geocoded SSS imagery rather than to recover the true physical attitude. Therefore, the refined pitch and yaw may mix contributions from multiple extrinsic error sources, which is acceptable provided that the refined imagery achieves improved geometric consistency.

\subsubsection{Port-Starboard Symmetry of Yaw and Pitch}

As shown in Fig. \ref{theory}, dual-sided SSS provides paired port-starboard imaging geometries for each ping, yielding consistent symmetry patterns of along-track distortions:

\begin{itemize}

\item \textbf{Yaw-antisymmetric:} A heading deflection advances one side along-track while delaying the other, producing opposite-signed mirrored distortions.

\item \textbf{Pitch-symmetric:} An along-track stacking-scale disturbance produces similar stretching and compression or same-signed misalignment trends on both sides.

\end{itemize}

We exploit this property by forming symmetric and antisymmetric components from joint port-starboard observations, approximately decoupling the pitch and yaw channels and improving both observability and robustness of estimation.

\begin{figure*}[htbp]
	\centering
	\includegraphics[width=1\linewidth]{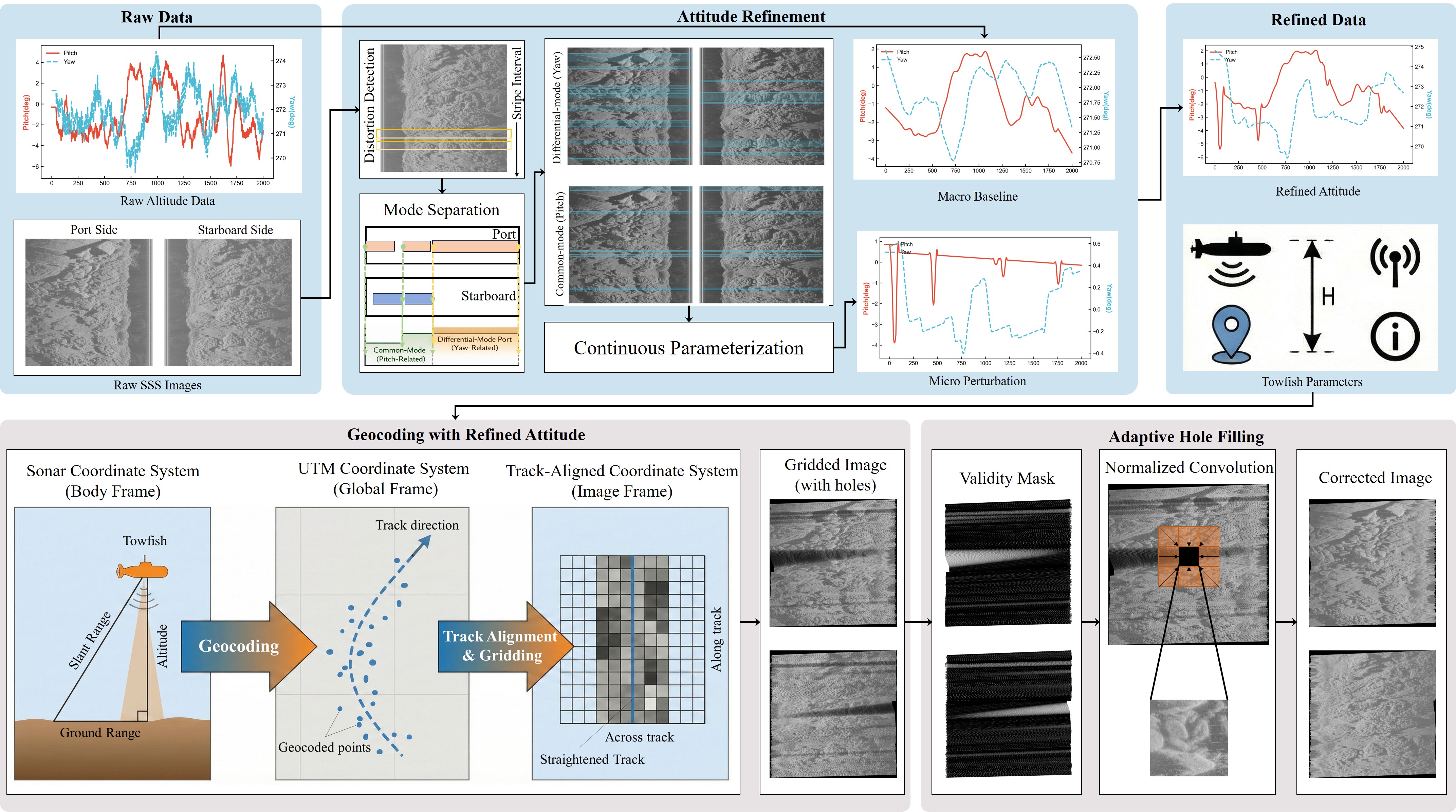}
	\caption{Overall pipeline of the proposed SSS geometric correction method. Given the raw yaw–pitch sequence and dual-sided waterfall images, the method first performs image-consistent attitude refinement by extracting stripe-wise distortion cues, separating them into differential-mode (yaw-related) and common-mode (pitch-related) components, converting them into continuous local perturbations, and combining them with a navigation-derived macro baseline to estimate the refined attitude. The refined attitude is then used for physically grounded geocoding and track-aligned gridding, and the gridding holes are completed by normalized convolution, yielding the final corrected SSS image.}
	\label{fig00}
\end{figure*}	

\subsection{Approach Overview}

Building on the equivalent yaw-pitch refinement model introduced above, the proposed method addresses SSS geometric correction through a unified framework that links image-consistent attitude refinement with refined-attitude geocoding, as illustrated in Fig. \ref{fig00}. The overall pipeline takes the raw attitude sequence and dual-sided waterfall images as input, and converts them into a geometrically corrected SSS image through the following two parts:

\begin{itemize}
    \item \textbf{Image-consistent attitude refinement:} A refined yaw-pitch sequence is estimated by combining a navigation-derived macro-scale baseline with image-inferred ping-scale perturbations. Specifically, stripe-wise distortion evidence is extracted from the port and starboard waterfall images through internal consistency analysis, organized into mode-separated supports, and converted into continuous local perturbations. These perturbations are then fused with the navigation baseline to obtain the refined attitude sequence.

    \item \textbf{Geocoding with refined attitude:} The refined attitude is incorporated into the geocoding stage for geometric correction. Each waterfall pixel is projected to its ground location and resampled onto a regular grid in a track-aligned metric frame, and the empty cells introduced during gridding are completed by normalized convolution using neighboring valid observations.
\end{itemize}

The resulting framework provides a complete path from raw sonar measurements to the corrected SSS image, improving local geometric consistency while preserving global geometric referencing.

\subsection{Navigation-Derived Attitude Baseline Estimation}

Consistent with the equivalent parameterization introduced above, we distinguish between the full physical attitude and the reduced attitude state used in the subsequent refinement. Let
\begin{equation}
\tilde{\boldsymbol{\Theta}}(n) = [\phi(n),\,\psi(n),\,\theta(n)]^T,
\end{equation}
denote a generic full attitude vector at ping index $n$, where $\phi$, $\psi$, and $\theta$ denote roll, yaw, and pitch, respectively. In particular, $\tilde{\boldsymbol{\Theta}}_{\mathrm{true}}(n)$ denotes the unknown physical ground-truth attitude, $\tilde{\boldsymbol{\Theta}}_{\mathrm{obs}}(n)$ denotes the navigation-observed attitude provided by onboard sensors, and $\tilde{\boldsymbol{\Theta}}_{\mathrm{ref}}(n)$ denotes the refined attitude used for geocoding. Here, $\tilde{\boldsymbol{\Theta}}_{\mathrm{true}}(n)$ is not directly available in practice, whereas $\tilde{\boldsymbol{\Theta}}_{\mathrm{obs}}(n)$ may be affected by noise, timing mismatch, and filtering artifacts. Under the modeling assumptions, roll is not treated as an explicit refinement variable because its contribution to the along-track geometric distortion considered here is secondary. Therefore, the subsequent derivation is conducted in the reduced yaw-pitch subspace,
\begin{equation}
\boldsymbol{\Theta}(n)=\Pi\bigl(\tilde{\boldsymbol{\Theta}}(n)\bigr)
=
[\psi(n),\,\theta(n)]^T,
\end{equation}
where $\Pi(\cdot)$ extracts the yaw and pitch components.

Accordingly, the reduced ground-truth attitude is
\begin{equation}
{{\bf{\Theta }}_{{\rm{true}}}}(n) = {[{\psi _{{\rm{true}}}}(n),\;{\theta _{{\rm{true}}}}(n)]^T},
\end{equation}
and the navigation provides the observed reduced attitude sequence
\begin{equation}
{{\bf{\Theta }}_{{\rm{obs}}}}(n) = {[{\psi _{{\rm{obs}}}}(n),\;{\theta _{{\rm{obs}}}}(n)]^T},
\end{equation}
which is typically affected by high-frequency noise, timing mismatch, and onboard filtering artifacts, leading to ping-scale discrepancies that can be amplified into visible striping and misalignment after geocoding.

To retain a reliable global geometric reference while enabling image-driven refinement at the line-scan scale, we decompose the refined attitude ${{\bf{\Theta }}_{{\rm{ref}}}}(n) = {[{\psi _{{\rm{ref}}}}(n),\;{\theta _{{\rm{ref}}}}(n)]^T}$ into a macro baseline and a micro perturbation:
\begin{equation}
\mathbf{\Theta}_{\mathrm{ref}}(n)=\bar{\mathbf{\Theta}}(n)+\delta\mathbf{\Theta}(n),
\end{equation}
where $\bar{\mathbf{\Theta}}(n)$ captures the low-frequency motion trend used as the global reference for geocoding, and $\delta\mathbf{\Theta}(n)$ models the ping-level residual to be estimated later from image consistency.

For the macro baseline, we estimate a smooth attitude trend $\bar{\boldsymbol{\Theta}}(n)=\left[\bar{\psi}(n),\,\bar{\theta}(n)\right]^T $ from the observed sequence $\boldsymbol{\Theta}_{\mathrm{obs}}(n)$ using a Savitzky-Golay (SG) smoother $\mathcal{H}_{\mathrm{SG}}(\cdot)$:
\begin{equation}
\bar{\boldsymbol{\Theta}}(n)
=
\mathcal{H}_{\mathrm{SG}}\!\left\{\boldsymbol{\Theta}_{\mathrm{obs}}(n)\right\}
=
\sum_{k=-K}^{K}\mu_k\,\boldsymbol{\Theta}_{\mathrm{obs}}(n+k),
\end{equation}
where $2K+1$ is the window length and $\{\mu_k\}$ are the SG coefficients. We use SG smoothing because it suppresses ping-scale jitter while better preserving the local low-order trend than simple moving-average filtering, which is desirable since the baseline is intended to retain the navigation-consistent macro motion rather than attenuate it excessively. In practice, $2K+1$ is chosen to be sufficiently larger than the typical span of stripe-level distortion events, so that $\bar{\boldsymbol{\Theta}}(n)$ captures only the low-frequency motion trend; in all experiments, $K$ is selected empirically according to the ping density and motion smoothness of the survey.

\subsection{Image-Based Microscopic Attitude Inference}

Given the navigation-derived baseline $\bar{\mathbf{\Theta}}(n)$, the remaining micro perturbation
$\delta\mathbf{\Theta}(n)=\big[\delta\psi(n),\ \delta\theta(n)\big]^T$
accounts for ping-scale stacking inconsistencies that manifest as along-track misalignment, as well as local stretching and compression, in waterfall imagery. Since these residual distortions are more directly observable in the image domain than in the navigation measurements, we infer $\delta\mathbf{\Theta}(n)$ from the distortion patterns in the waterfall image.

\subsubsection{Stripe-wise Distortion Detection}

To infer the micro perturbation from image, we detect residual distortions at the stripe level rather than at the individual-ping level. Ping-scale changes are often weak and nonstationary, but over a short along-track span they exhibit more coherent structural patterns. We therefore aggregate rows into overlapping stripes to obtain more stable distortion evidence. Within each stripe, we use multi-scale ROIs rather than the full stripe as a whole, because the distortion response may vary across the swath. In particular, yaw-induced deformation is range-dependent, so its along-track distortion is typically stronger at some across-track locations than at others. ROI-based evaluation preserves such localized cues, and the multi-scale design improves robustness to distortions of different spatial extents. The resulting ROI responses are then aggregated into a stripe-wise score sequence for subsequent distortion detection.

\paragraph{Overlapped stripe construction}

Let $I\in\mathbb{R}^{W\times H}$ denote the SSS waterfall image, where column index $x\in\{0,\ldots,W-1\}$ corresponds to across-track and row index $y\in\{0,\ldots,H-1\}$ corresponds to along-track. With stripe height $h_s$ and along-track stride $\Delta_y<h_s$ (to enforce overlap and reduce incidental detections in texture-sparse or scattering-unstable regions), the $j$-th stripe is
\begin{equation}
T_j = I[0:W-1,\ y_j:y_j+h_s-1], \quad y_j=j\Delta_y,
\end{equation}
yielding the stripe set $\mathcal{T}=\{T_j\}_{j=0}^{M-1}$, where $M$ represents the total number of stripes.

\paragraph{Multi-scale ROIs (Region of Interest)}

Within each stripe $T_j$, we form a scale set ${\cal S} = \{ {s_1}, \ldots ,{s_{{N_s}}}\} \subset (0,1)$ and an overlap ratio $\rho\in(0,1)$. For a given scale $s \in \mathcal{S}$, the ROI width is $sW$ and stride is $\Delta_x(s)=(1-\rho)sW$. So the $i$-th ROI at scale $s$ in stripe $j$ is defined as:
\begin{equation}
r_{j,s,i}=I[\ x_{s,i}:x_{s,i}+sW-1,\ y_j:y_j+h_s-1], x_{s,i}=i\Delta_x(s),
\end{equation}
and the ROI set is $\mathcal{R}_j(s)=\{r_{j,s,i}\}_{i=0}^{M_s-1}$, where $M_s$ is the total number of ROIs at scale $s$ in stripe $j$. Larger ROIs provide stable structural statistics, while smaller ROIs respond to localized geometric peaks.

\paragraph{Consistency measurement}
\begin{figure*}[htbp]
	\centering
	\includegraphics[width=1\linewidth]{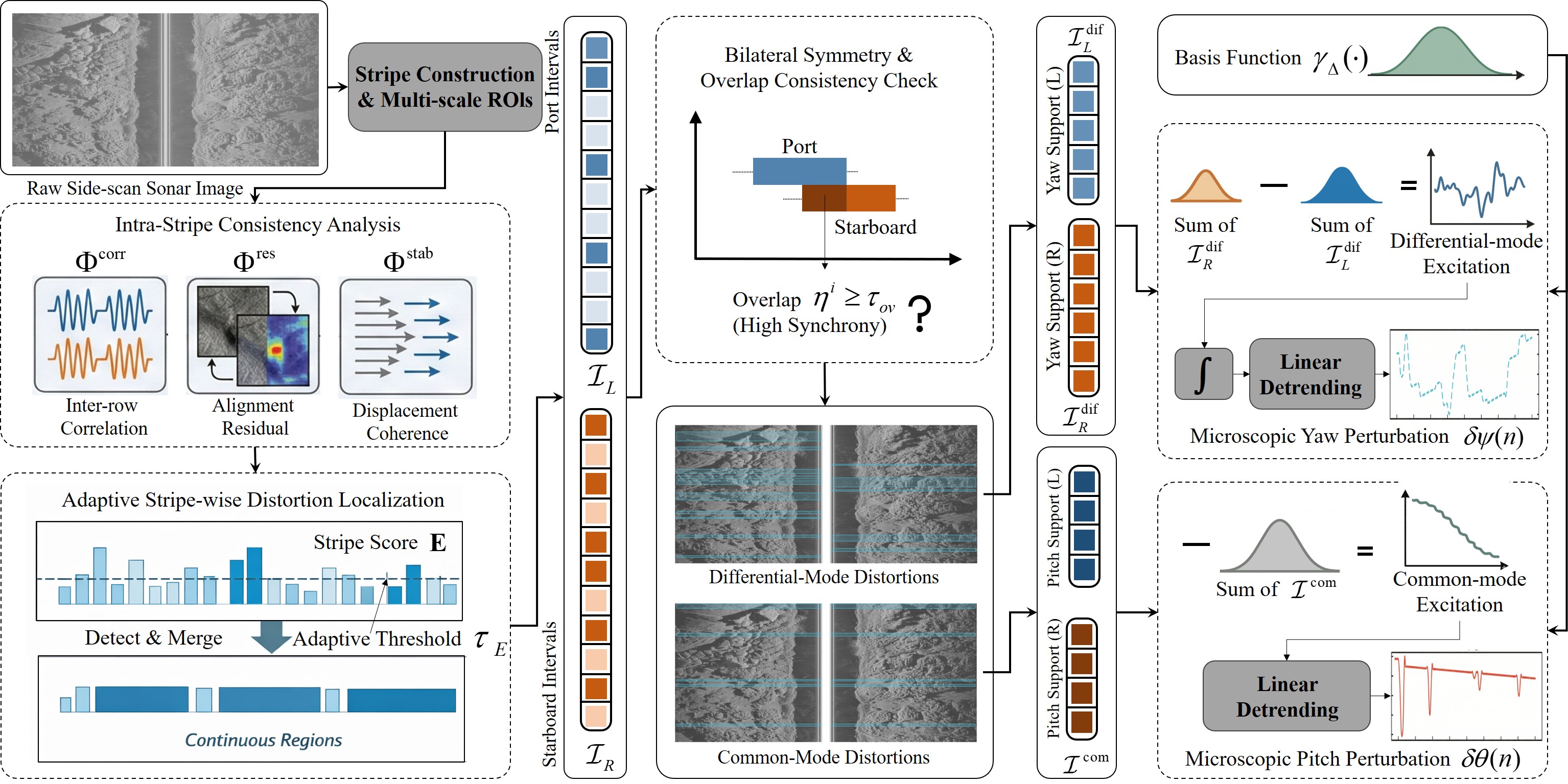}
	\caption{Flowchart of the proposed image-based microscopic attitude inference. Overlapping stripes and multi-scale ROIs are first constructed on the raw SSS waterfall image, and stripe-wise distortion evidence is extracted using three intra-stripe consistency measures: correlation enhancement, post-alignment residual, and displacement coherence. After adaptive thresholding and interval merging, the detected port and starboard distortion intervals are separated by cross-swath overlap consistency into differential-mode supports (yaw-related) and common-mode support (pitch-related). These supports are then converted into smooth basis excitations, from which the microscopic yaw perturbation $\delta\psi(n)$ and pitch perturbation $\delta\theta(n)$ are generated through sign-aware combination, integration, and linear detrending.}
	\label{fig01}
\end{figure*}	

Based on the overlapped stripes and multi-scale ROIs defined above, we quantify local image consistency to identify distortion patterns useful for attitude refinement. Under the reduced yaw-pitch formulation, the dominant residual distortions mainly appear as local stretching and compression of adjacent scan lines. In this work, we focus on stretching only, because it yields a more verifiable image signature. Specifically, local stretching introduces redundant sampling, so adjacent rows within the affected ROI tend to become more similar. Moreover, this similarity is not purely appearance-based, it can usually be explained by a small horizontal realignment of the across-track profiles, and the required offsets remain relatively consistent across neighboring rows. Compression, in contrast, mainly causes information crowding and reduced separability, whose appearance is easily confounded with natural seabed texture variation. As shown in Fig. \ref{fig01}, we therefore characterize each ROI by three complementary indicators that correspond to these three observable properties of stretching: correlation response $\Phi^{\mathrm{corr}}$, post-alignment residual $\Phi^{\mathrm{res}}$, and displacement coherence $\Phi^{\mathrm{stab}}$.

Specifically, for an ROI $r_{j,s,i}$, we define
\begin{equation}
\mathbf{f}_y = I[x_{s,i}:x_{s,i}+sW-1,\ y],
y\in\{y_j,\ldots,y_j+h_s-1\},
\end{equation}
where $\mathbf{f}_y$ denotes the 1-D across-track profile extracted from row $y$ within the ROI. Let $y_m = y_j + h_s/2$ be the reference-row index, and take $\mathbf{f}_{y_m}$ as the reference profile. The set of all non-center rows is then
$\mathcal{Y}=\{y_j,\ldots,y_j+h_s-1\}\backslash\{y_m\}$.
For each $\mathbf{f}_y$ with $y\in\mathcal{Y}$, we estimate, by 1-D phase correlation, the horizontal offset $\Delta_y$ along the $x$-axis that best aligns $\mathbf{f}_y$ to $\mathbf{f}_{y_m}$, together with the corresponding peak correlation response $\gamma_y\in[0,1]$. Accordingly, $\Gamma(\mathbf{f}_y,\Delta_y)$ denotes the shift-compensated profile of row $y$ after alignment to the reference row. Based on these quantities, we define three complementary metrics,
\begin{equation}
\begin{aligned}
\Phi_{j,s,i}^{\mathrm{corr}} &= \frac{1}{|\mathcal{Y}|} \sum_{y \in \mathcal{Y}} \gamma_y, \\
\Phi_{j,s,i}^{\mathrm{res}} &= \frac{1}{|\mathcal{Y}|} \sum_{y \in \mathcal{Y}} 
\frac{\left\| \mathbf{f}_{y_m} - \Gamma(\mathbf{f}_y, \Delta_y) \right\|_1}
{\left\| \mathbf{f}_{y_m} \right\|_1 + \varepsilon}, \\
\Phi_{j,s,i}^{\mathrm{stab}} &= \frac{1}{|\mathcal{Y}|} \sum_{y \in \mathcal{Y}} 
\left| \Delta_y - \tilde{\Delta} \right|,
\tilde{\Delta} = \mathcal{F}_{\mathrm{med}}\!\left( \{ \Delta_y \}_{y \in \mathcal{Y}} \right),
\end{aligned}
\end{equation}
where $\varepsilon$ is a small positive stabilizer, and $\mathcal{F}_{\mathrm{med}}(\cdot)$ denotes the median operator. Here, $\Phi^{\mathrm{corr}}$ measures the row-wise similarity within the ROI, but high similarity alone is not sufficient evidence of stretching, since repetitive or weakly textured seabed regions may also produce similar profiles. To suppress such false responses, $\Phi^{\mathrm{res}}$ evaluates whether the similarity can be effectively removed by horizontal shift compensation. A genuine stretching region is expected to yield a low post-alignment residual because adjacent rows become redundant observations of the same local structure. In addition, $\Phi^{\mathrm{stab}}$ evaluates whether the estimated offsets remain consistent across rows. This is because a local stretching event affects neighboring scan lines in a continuous manner, whereas noisy or texture-driven matches usually produce irregular offsets. Accordingly, a genuine stretching region should exhibit not only high $\Phi^{\mathrm{corr}}$, but also low $\Phi^{\mathrm{res}}$ and low $\Phi^{\mathrm{stab}}$.

To aggregate these cues into a single internal-consistency score, we define
\begin{equation}
E_{j,s,i}=\Phi^{\mathrm{corr}}_{j,s,i}-\lambda\,\Phi^{\mathrm{res}}_{j,s,i}-\mu\,\Phi^{\mathrm{stab}}_{j,s,i},
\end{equation}
so that $E_{j,s,i}$ becomes high only when the ROI simultaneously exhibits strong row similarity, good explainability by horizontal realignment, and coherent offsets across neighboring rows

To obtain a stripe-level distortion measure, we aggregate the ROI-level internal-consistency scores $\{E_{j,s,i}\}$ associated with stripe $j$. Here, the goal is not to measure the average distortion strength over the entire stripe, but to determine whether the stripe contains a locally salient stretching cue. In practice, residual stretching is often spatially localized and may be prominent only within a limited across-track region and at a particular ROI scale. Under such conditions, averaging would favor spatial prevalence rather than distortion saliency, and may therefore suppress precisely the localized events we aim to detect. We therefore use max-pooling over ROI locations and scales to preserve the strongest distortion evidence within the stripe:
\begin{equation}
E_j=\max_{s\in\mathcal{S}}\ \max_i\ E_{j,s,i},\quad
\mathbf{E}=[E_0,\ldots,E_{M-1}]^{\top},
\end{equation}
where $E_j$ denotes the final distortion score of the $j$-th stripe, and $\mathbf{E}$ forms the 1-D stripe-wise evidence sequence for subsequent distortion localization.

\paragraph{Adaptive Stripe-wise Distortion Localization}

Given the stripe-wise evidence sequence $\mathbf{E}$, distorted stripes are identified by adaptive thresholding. Since $\{E_j\}$ may contain outlying high responses induced by terrain relief and scattering nonstationarity, mean-variance statistics are not sufficiently robust. We therefore adopt a median-centered threshold,
\begin{equation}
\tau_E=\tilde{E}+\kappa \cdot \frac{1}{M}\sum_{j=0}^{M-1}\left|E_j-\tilde{E}\right|,
\quad
\tilde{E}=\mathcal{F}_{\mathrm{med}}\!\left(\{E_j\}\right),
\end{equation}
where $\kappa>0$ controls the detection strictness. The stripe-level decision is then given by
\begin{equation}
b_j=
\begin{cases}
1, & E_j>\tau_E,\\
0, & \text{otherwise}.
\end{cases}
\end{equation}

Because physically meaningful distortion events are expected to be continuous along track, the binary decisions $\{b_j\}$ are further regularized in the row domain. Each detected stripe ($b_j=1$) is first mapped to its covered row interval $[y_j,\ y_j+h_s-1]$. Adjacent or overlapping intervals are then merged, and short isolated detections are removed using a minimum-length constraint. This yields the final set of detected distortion intervals,
\begin{equation}
{\cal I} = \{ {\Delta ^m} = [y _{{\rm{st}}}^m,y _{{\rm{ed}}}^m]\} _{m = 1}^{{M_I}},
\end{equation}
where $(y_{\mathrm{st}}^m,y_{\mathrm{ed}}^m)$ denotes the start and end row indices of the $m$-th detected interval.

\subsubsection{Port-Starboard Mode Separation}

The stripe-wise distortion localization described above is applied independently to the port and starboard waterfall images, yielding two sets of detected along-track intervals $\mathcal{I}_L=\{\Delta_L^i\}_{i=1}^{M_L}$ and $\mathcal{I}_R=\{\Delta_R^j\}_{j=1}^{M_R}$, where each interval represents a detected stretching event on the corresponding swath. Under the port-starboard symmetry, pitch-induced disturbances tend to produce stretching responses on both swaths simultaneously, whereas yaw-induced disturbances tend to produce stretching on only one swath and compression on the other. Since the proposed detector is sensitive to stretching but not compression, pitch-related events are expected to appear as synchronized detections on the two swaths, while yaw-related events are expected to appear as one-sided detections.

Based on this observation, as shown in Fig. \ref{fig01}, we separate the detected intervals into common-mode and differential-mode supports according to their cross-swath overlap. For an interval pair $(\Delta_L^i, \Delta_R^j)$, the overlap ratio is defined as
\begin{equation}
\eta_{ij}=\frac{|\Delta_L^i\cap \Delta_R^j|}{\min\!\left(|\Delta_L^i|,|\Delta_R^j|\right)},
\end{equation}
where $|\cdot|$ denotes the interval length. Using the shorter interval in the denominator makes $\eta_{ij}$ a coverage-oriented measure, which is more suitable than a duration-sensitive similarity measure because synchronized detections on the two swaths may have unequal extents. For each detected interval, its maximum overlap ratio with all intervals on the opposite swath is defined as
\begin{equation}
\eta_L^i=\max_{1\le j\le M_R}\eta_{ij},\quad
\eta_R^j=\max_{1\le i\le M_L}\eta_{ij}.
\end{equation}

An interval is regarded as synchronized if its maximum overlap ratio exceeds a threshold ${\tau_{ov}}\in(0,1]$; otherwise, it is treated as a one-sided detection. Accordingly, all synchronized detections from the two swaths are collected and merged to form the common-mode support $\mathcal{I}^{\mathrm{com}}$, while the remaining unmatched detections define the differential-mode supports
\begin{equation}
\mathcal{I}^{\mathrm{dif}}_L=\{\Delta_L^i:\eta_L^i<\tau_{ov}\},\quad
\mathcal{I}^{\mathrm{dif}}_R=\{\Delta_R^j:\eta_R^j<\tau_{ov}\},
\end{equation}
here, $\mathcal{I}^{\mathrm{com}}$ represents regions of synchronized stretching primarily associated with pitch-related disturbances, whereas $\mathcal{I}^{\mathrm{dif}}_L$ and $\mathcal{I}^{\mathrm{dif}}_R$ represent one-sided stretching regions typically associated with yaw-related disturbances.

\subsubsection{Support-Driven Continuous Parameterization}

After obtaining the common-mode and differential-mode supports,
$\mathcal{I}^{\mathrm{com}}$, $\mathcal{I}^{\mathrm{dif}}_L$, and $\mathcal{I}^{\mathrm{dif}}_R$,
the remaining task is to convert these detected distortion regions into continuous microscopic attitude perturbations $\delta\mathbf{\Theta}(n)=[\delta\psi(n),\,\delta\theta(n)]^T$.
Since attitude disturbances affect neighboring pings continuously rather than independently, as shown in Fig. \ref{fig01}, we model each detected interval as a smooth along-track excitation and then construct $\delta\psi(n)$ and $\delta\theta(n)$ according to the imaging roles of yaw and pitch.

For an interval $\Delta  = [{y_{{\rm{st}}}},{y_{{\rm{ed}}}}]$, we define a smooth finite-support basis function
\begin{equation}
{\gamma_\Delta }(n)=
\begin{cases}
\beta \!\left(\dfrac{n-{y_{{\rm{st}}}}}{{y_{{\rm{ed}}}}-{y_{{\rm{st}}}}}\right), & n\in[{y_{{\rm{st}}}},{y_{{\rm{ed}}}}],\\[4pt]
0, & \text{otherwise},
\end{cases}
\end{equation}
where $\beta(t)=\frac{1}{2}\!\left(1-\cos(2\pi t)\right)$, $t\in[0,1]$, is the normalized Hann window \cite{A21}. This basis provides a smooth onset and offset for each detected interval and avoids unrealistically abrupt attitude changes at interval boundaries.

The yaw perturbation is constructed from the differential-mode supports. Under the adopted port-starboard sign convention, a one-sided stretching detection on the starboard side indicates a positive yaw contribution, whereas a one-sided stretching detection on the port side indicates a negative yaw contribution. We therefore define the differential-mode excitation as
\begin{equation}
d_\psi(n)=\sum_{\Delta\in\mathcal{I}^{\mathrm{dif}}_R}{\gamma _\Delta }(n)
-\sum_{\Delta\in\mathcal{I}^{\mathrm{dif}}_L}{\gamma _\Delta }(n).
\end{equation}

Because yaw affects the image through the accumulated heading variation along track \cite{A22}, the microscopic yaw perturbation is obtained by discrete integration,
\begin{equation}
\delta\psi(n)=\sum_{\tau=1}^{n}d_\psi(\tau), \quad n=0,\ldots,H-1.
\end{equation}

The pitch perturbation is constructed from the common-mode support. Since pitch-related distortion appears as synchronized stretching on the two swaths, its support is directly provided by $\mathcal{I}^{\mathrm{com}}$. Under the stretching-only convention adopted here, stretching corresponds to a decrease in pitch angle. We therefore define the common-mode excitation as
\begin{equation}
d_\theta(n)=-\sum_{\Delta\in\mathcal{I}^{\mathrm{com}}}{\gamma _\Delta }(n),
\end{equation}
and use it directly as the microscopic pitch perturbation,
\begin{equation}
\delta\theta(n)=d_\theta(n), \quad n=0,\ldots,H-1.
\end{equation}

Finally, $\delta\psi(n)$ and $\delta\theta(n)$ are each smoothed and linearly detrended \cite{A23} to suppress residual discontinuities and long-term bias, yielding the final microscopic attitude estimates used in the subsequent correction stage.

\subsection{Fusion of Baseline and Microscopic Components}

The preceding steps provide two complementary attitude components: the navigation-derived baseline
$\bar{\mathbf{\Theta}}(n)=[\bar{\psi}(n),\,\bar{\theta}(n)]^T$,
which captures the reliable low-frequency motion trend, and the image-derived microscopic perturbation
$\delta\mathbf{\Theta}(n)=[\delta\psi(n),\,\delta\theta(n)]^T$,
which describes the residual line-scan-scale disturbance inferred from distortion evidence. Given that these two components represent distinct scales of the same platform attitude, the microscopic component has been smoothed and detrended as a local correction term, allowing the refined attitude sequence to be obtained by direct superposition:
\begin{equation}
\psi_{\mathrm{ref}}(n)=\bar{\psi}(n)+\delta\psi(n),\quad
\theta_{\mathrm{ref}}(n)=\bar{\theta}(n)+\delta\theta(n),
\end{equation}
here, the baseline component preserves the global navigation-consistent trend, while the microscopic component compensates for localized distortion-induced attitude deviations. The fused refined sequence ${{\bf{\Theta }}_{{\rm{ref}}}}(n) = {[{\psi _{{\rm{ref}}}}(n),\;{\theta _{{\rm{ref}}}}(n)]^T}$ is then used as the attitude input for the subsequent geometric correction.

\subsection{Geocoding with Refined Attitude}

\begin{figure*}[htbp]
	\centering
	\includegraphics[width=1\linewidth]{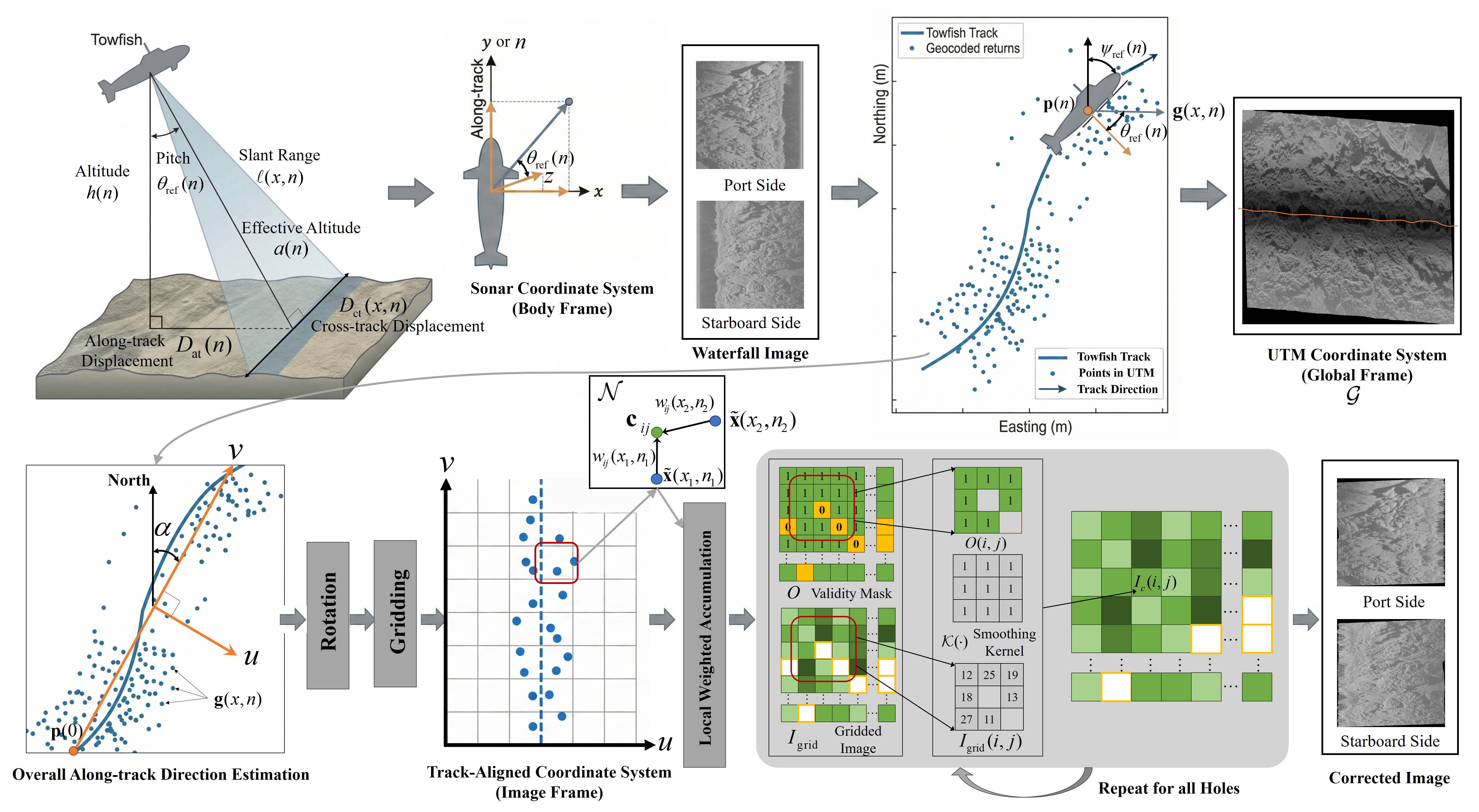}
	\caption{Illustration of the refined-attitude geocoding process. Starting from the waterfall image coordinates, each pixel is assigned a ground-plane offset according to slant range, altitude, and refined pitch, and is then rotated by the refined yaw and translated by the towfish position to obtain its location in the global UTM frame. The resulting irregular geo-referenced samples are further rotated into a track-aligned local metric frame for image formation, where regular gridding is performed by local weighted accumulation. A validity mask is generated simultaneously to indicate empty cells, and normalized convolution is applied to fill these holes from neighboring valid samples, yielding the final geometrically corrected SSS image.}
	\label{fig02}
\end{figure*}

After fusion, as shown in Fig. \ref{fig02}, the refined attitude sequence $ \boldsymbol{\Theta}_{\mathrm{ref}}(n), n=0,\ldots,H-1,$ is taken as the attitude input to the final geocoding stage. At this stage, the problem is no longer attitude inference; instead, the goal is to map each waterfall pixel to its physical ground location using the refined pose parameters, and then resample the resulting irregularly distributed geo-referenced samples into a regular image representation. In addition to the global Universal Transverse Mercator (UTM) coordinate system \cite{A24} used for physical localization, we further introduce a track-aligned local metric frame to facilitate image formation. The motivation is that the survey trajectory is generally not aligned with the axes of the global map projection, so direct rasterization in UTM coordinates would produce an output grid with an inconvenient orientation and potentially more empty cells. By defining a local frame whose axes are aligned with the dominant survey direction, the final image can preserve the natural across-track/along-track organization of the sonar acquisition while remaining metrically consistent with the global coordinates.

To keep the formulation consistent with the preceding sections, we explicitly distinguish the three coordinate systems used in this subsection:
\begin{itemize}
    \item the waterfall image coordinates $(x,y)$, where $x$ and $y$ denote the across-track and along-track pixel indices, respectively;
    
    \item the global UTM coordinates $\mathbf{p}=[p_x,p_y]^T$, used to represent physical ground locations;
    
    \item the track-aligned local metric frame $(u,v)$, whose axes are defined, respectively, along the across-track and along-track directions of the survey path, and which is used as the regular output grid for resampling.
\end{itemize}

\subsubsection{Pixel-Wise Ground Projection in the UTM Frame}

Given the refined attitude sequence
$\boldsymbol{\Theta}_{\mathrm{ref}}(n)
$,
the first step of geocoding is to assign each single-side waterfall pixel to its physical ground location in the UTM frame. Note that this subsection switches from the image-coordinate notation $I(x,y)$ used above to the equivalent notation $I(x,n)$. This is only a change of indexing, not of the image representation itself. Because the single-side waterfall image is formed by sequentially stacking ping returns, each row index $y$ corresponds exactly to one ping index $n$, i.e., $y=n$. Hence, $x\in\{0,\ldots,W-1\}$ denotes the across-track sample index, while $n\in\{0,\ldots,H-1\}$ denotes the ping index, equivalently the row index of the waterfall image.

For ping $n$, let $\mathbf{p}(n)=[p_x(n),\,p_y(n)]^T$ denote the towfish position in the UTM frame, $h(n)$ the altitude, and $\ell_{\max}(n)$ the maximum slant range. The mapping is performed by first expressing each sample as a local ground-plane offset relative to the sonar, and then transforming this offset into the global UTM frame using the refined yaw angle. Because the waterfall image is sampled in slant range, the sample at pixel $(x,n)$ is first associated with the slant-range distance
\begin{equation}
\ell(x,n)=\ell_{\max}(n)\frac{x}{W-1}.
\end{equation}

Following the reduced yaw-pitch model used in the previous subsections, the refined pitch angle affects the projection from slant range to the ground plane through the effective altitude
\begin{equation}
a(n)=\frac{h(n)}{\cos\theta_{\mathrm{ref}}(n)}.
\end{equation}

The corresponding ground-range distance is then
\begin{equation}
D(x,n)=\sqrt{\ell(x,n)^2-a(n)^2}.
\end{equation}

Let $\sigma\in\{-1,+1\}$ denote the swath-side indicator, with $\sigma=-1$ for port and $\sigma=+1$ for starboard. Then the local cross-track coordinate of the projected sample is
\begin{equation}
D_{\mathrm{ct}}(x,n)=\sigma\,D(x,n).
\end{equation}

Under the same reduced model, the refined pitch also induces an along-track displacement of the projected footprint, given by
\begin{equation}
D_{\mathrm{at}}(n)=h(n)\tan\theta_{\mathrm{ref}}(n).
\end{equation}

Accordingly, the projected location of pixel $(x,n)$ in the local sonar-ground frame is written as
\begin{equation}
\mathbf{D}_{\mathrm{loc}}(x,n)=
\begin{bmatrix}
D_{\mathrm{ct}}(x,n)\\
D_{\mathrm{at}}(n)
\end{bmatrix}.
\end{equation}

This local ground-plane offset is then rotated by the refined yaw angle and translated by the towfish position to obtain its UTM coordinates:
\begin{equation}
    {\bf{g}}(x,n) = {\bf{p}}(n) + \left[ {\begin{array}{*{20}{c}}
    {\cos {\psi _{{\rm{ref}}}}(n)}&{ - \sin {\psi _{{\rm{ref}}}}(n)}\\
    {\sin {\psi _{{\rm{ref}}}}(n)}&{\cos {\psi _{{\rm{ref}}}}(n)}
    \end{array}} \right]{{\bf{D}}_{{\rm{loc}}}}(x,n).
\end{equation}

Applying this mapping to all valid pixels converts the waterfall image into an irregular set of geo-referenced samples,
\begin{equation}
\mathcal{G}
=
\left\{
\bigl(\mathbf{g}(x,n),\,I(x,n)\bigr)
\;\middle|\;
x=0,\ldots,W-1,
n=0,\ldots,H-1
\right\},
\end{equation}
where $\mathbf{g}(x,n)\in\mathbb{R}^2$ is the UTM location associated with pixel $I(x,n)$. This irregular sample set serves as the input to the subsequent gridding step.

\subsubsection{Gridding in the Track-Aligned Local Frame}

After pixel-wise projection, as shown in Fig. \ref{fig02}, the waterfall image is converted into the irregular geo-referenced sample set $\mathcal{G}$. The next step is to resample these irregular samples onto a regular image grid in the track-aligned local metric frame. To this end, we determine an overall along-track direction from the towfish trajectory. This definition assumes that the considered survey segment has a well-defined dominant motion direction; if strong turning or return motion is present, the start-end displacement may no longer provide a stable orientation estimate, and the trajectory should be segmented before gridding. Using the start and end positions, $\mathbf{p}(0)$ and $\mathbf{p}(H-1)$, the unit along-track direction is defined as
\begin{equation}
\mathbf{t}
=
\frac{\mathbf{p}(H-1)-\mathbf{p}(0)}
{\|\mathbf{p}(H-1)-\mathbf{p}(0)\|_2}
=
[t_x,\,t_y]^T,
\end{equation}
with azimuth $\alpha=\operatorname{atan2}(t_y,t_x)$. The corresponding across-track unit vector is
\begin{equation}
\mathbf{n}
=
[-\sin\alpha,\ \cos\alpha]^T.
\end{equation}

Using these two orthogonal unit vectors, we define a track-aligned local metric frame, in which $\mathbf{n}$ and $\mathbf{t}$ specify the across-track and along-track basis directions, respectively. The coordinates of each projected sample in this frame are denoted by $(u,v)$. Taking $\mathbf{p}(0)$ as the origin of this local frame, each projected sample is transformed from the UTM frame to the local metric frame by
\begin{equation}
\mathbf{q}(x,n)=
\begin{bmatrix}
u(x,n)\\
v(x,n)
\end{bmatrix}
=
\begin{bmatrix}
\mathbf{n}^T\\
\mathbf{t}^T
\end{bmatrix}
\bigl(\mathbf{g}(x,n)-\mathbf{p}(0)\bigr).
\end{equation}

A regular output grid of size $W\times H$ is then defined over the bounding box of the transformed samples in the $(u,v)$ plane. Each transformed sample ${\bf{\tilde x}}(x,n) = {\left[ {\tilde x(x,n),\tilde y(x,n)} \right]^T}$ is linearly mapped to continuous image coordinates on this grid as
\begin{equation}
\left[ {\begin{array}{*{20}{c}}
{\tilde x(x,n)}\\
{\tilde y(x,n)}
\end{array}} \right] = \left[ {\begin{array}{*{20}{c}}
{\frac{{W - 1}}{{{u_{\max }} - {u_{\min }}}}}&0\\
{0}&{\frac{{H - 1}}{{{v_{\max }} - {v_{\min }}}}}
\end{array}} \right]\left[ {\begin{array}{*{20}{c}}
{u(x,n) - {u_{\min }}}\\
{v(x,n) - {v_{\min }}}
\end{array}} \right],
\end{equation}
where $u_{\min},u_{\max},v_{\min}$, and $v_{\max}$ are determined from the extrema of the transformed sample coordinates.

Because the mapped sample locations are generally non-integer, multiple samples may contribute to the same output pixel, while some pixels may receive no samples. We therefore compute the gridded image by local weighted accumulation. For the output pixel $(i,j)$ with center $\mathbf{c}_{ij}=[i,\,j]^T$, the contribution of sample $(x,n)$ is weighted by
\begin{equation}
    {w_{ij}}(x,n) = \frac{1}{{\left\| {{{\left[ {\tilde x(x,n),\tilde y(x,n)} \right]}^T} - {{\bf{c}}_{ij}}} \right\|_2^2 + \varepsilon }},
\end{equation}
where $\varepsilon>0$ is a small stabilizing constant. Denoting by $\mathcal{N}(i,j)$ the set of projected samples contributing to pixel $(i,j)$, the gridded image ${I_{{\rm{grid}}}}$ and the corresponding validity mask $O$ are given by
\begin{equation}
\begin{array}{l}
\displaystyle
I_{{\rm grid}}(i,j)=
\frac{\sum\nolimits_{(x,n)\in\mathcal N(i,j)} w_{ij}(x,n)I(x,n)}
{\sum\nolimits_{(x,n)\in\mathcal N(i,j)} w_{ij}(x,n)},\\
O(i,j)=\mathbf{1}\left(
\sum\nolimits_{(x,n)\in\mathcal N(i,j)} w_{ij}(x,n)>0
\right).
\end{array}
\end{equation}

\subsubsection{Grid Completion by Normalized Convolution}

As described in the previous step, forward accumulation of irregular projected samples onto the regular output grid yields both a gridded image $I_{\mathrm{grid}}(i,j)$ and a validity mask $O(i,j)$. Pixels with $O(i,j)=0$ correspond to output cells that receive no projected samples, i.e., empty grid cells generated during the resampling process. To improve the spatial continuity of the final geocoded image while preserving the support of valid measurements, as shown in Fig. \ref{fig02}, these empty cells are completed using normalized convolution \cite{A25} from neighboring valid pixels only.

Specifically, for an output pixel $(i,j)$, let $\Lambda(i,j)$ denote a local window centered at $(i,j)$, and let ${\cal K}( \cdot )$ be a normalized convolution kernel defined on this window. The completed corrected image ${I}_c(i,j)$ is computed as
\begin{equation}
{I}_c(i,j)
=
\frac{
\sum\limits_{({i'},{j'})\in {\Lambda}(i,j)}
\mathcal{K}(i-{i'},j-{j'})\,O({i'},{j'})\,I_{\mathrm{grid}}({i'},{j'})
}{
\sum\limits_{({i'},{j'})\in{\Lambda}(i,j)}
\mathcal{K}(i-{i'},j-{j'})\,O({i'},{j'})+\varepsilon
},
\end{equation}
where $I_{\mathrm{grid}}$ is the input signal, $O$ acts as a binary confidence map, and the convolution kernel $\mathcal{K}$ controls the spatial weighting of neighboring pixels. Because the mask explicitly restricts the support to valid grid cells, only pixels with $O({i'},{j'})=1$ contribute to the estimate. 

In this way, the refined attitude sequence is translated into a geometrically corrected SSS image on a regular grid through projection, track-aligned resampling, and normalized-convolution-based grid completion. The resulting image preserves global georeferencing through the navigation information while improving local geometric consistency through the refined yaw and pitch estimates.

\section{Experiment}

\begin{table*}[t]
\centering
\caption{Summary of the three side-scan sonar datasets used in this paper.}
\label{datasets}
\renewcommand{\arraystretch}{1.2} % Increase row height for readability
\setlength{\tabcolsep}{15.5pt}       % Adjusted to a normal spacing

\small
\begin{tabular}{lccc}
\toprule
\textbf{Dataset} & \textbf{I} & \textbf{II} & \textbf{III} \\
\midrule
Environment / Site & Sector N07 \cite{A26}  & Sector N08 \cite{A26} & Sector N07 \cite{A26} \\
Sonar Model & Marine Sonic Arc Scout MK II & Marine Sonic Arc Scout MK II & Klein 3000H \\
Operating Frequency & 900 kHz & 900 kHz & 500 kHz \\
Sensor Range & 30-60 m & 30-80 m & 50-100 m \\
Acquisition Altitude & $\sim$10\% of range &  $\sim$10\% of range &  $\sim$10\% of range \\
Precise Navigation Data & Yes & Yes & No \\
Geocoding available? & Yes & Yes & No \\
Role & Evaluation & Evaluation & Generalization Test \\
\bottomrule
\end{tabular}
\end{table*}

\subsection{Data Description}

Experiments are conducted on three SSS datasets acquired under different environmental conditions and sonar configurations, whose detailed acquisition parameters are summarized in Table \ref{datasets}. Among them, Dataset I and II are used for quantitative evaluation because they are accompanied by high-accuracy navigation and attitude records, which enable reliable geocoding of the raw sonar observations and thus provide reference images for objective comparison. By contrast, Datasets III does not support equally reliable reference geocoding due to the lack of sufficiently accurate auxiliary measurements. They are therefore used mainly for assessing the generalization of the proposed method across different scenes and sonar platforms.

\subsection{Evaluation Metrics}

For quantitative evaluation on Dataset I, let $I_r$ and $I_c$ denote the reference image and the corrected result, respectively, both normalized to $[0,1]$. To assess correction quality from complementary aspects, we report PSNR, MS-SSIM, NCC, and LPIPS.

\subsubsection{Peak Signal-to-Noise Ratio (PSNR)}

PSNR is computed from the mean squared error between $I_c$ and $I_r$:
\begin{equation}
{\rm{PSNR}} = 10{\log _{10}}\left( {\frac{1}{\varsigma }} \right),\varsigma  = \frac{1}{{|\Omega |}}\sum\limits_{x \in \Omega } {{{\left( {{I_c}(x) - {I_r}(x)} \right)}^2}} 
\end{equation}
where $\Omega$ denotes the image domain. A higher PSNR indicates a smaller pixel-wise reconstruction error.

\subsubsection{Multi-Scale Structural Similarity (MS-SSIM)}

MS-SSIM \cite{A27} measures structural agreement between $I_c$ and $I_r$ over multiple resolutions:
\begin{equation}
\mathrm{MS\text{-}SSIM}(I_c,I_r)
=
l_M^{\alpha_M}
\prod_{j=1}^{M}
c_j^{\beta_j}s_j^{\gamma_j},
\end{equation}
where $l_M$, $c_j$, and $s_j$ denote the luminance, contrast, and structure comparison terms defined in the standard MS-SSIM formulation at the corresponding scales, and $\alpha_M$, $\beta_j$, and $\gamma_j$ are the associated weights. Larger values indicate better structural consistency.

\subsubsection{Normalized Cross-Correlation (NCC)}

NCC evaluates the global correlation between the corrected image and the reference image:
\begin{equation}
\mathrm{NCC}
=
\frac{
\left\langle I_r-\mu_r,\; I_c-\mu_c \right\rangle
}{
\|I_r-\mu_r\|_2\,\|I_c-\mu_c\|_2
},
\end{equation}
where $\mu_r$ and $\mu_c$ are the mean intensities of $I_r$ and $I_c$, respectively, and $\langle \cdot,\cdot\rangle$ denotes the inner product over $\Omega$. A value closer to $1$ indicates stronger statistical consistency.

\subsubsection{Learned Perceptual Image Patch Similarity (LPIPS)}

LPIPS \cite{A28} measures perceptual similarity in a pretrained deep feature space:
\begin{equation}
\mathrm{LPIPS}(I_c,I_r)
=
\sum_{l}
w_l\,
\frac{1}{H_lW_l}
\left\|
\phi_c^{\,l}-\phi_r^{\,l}
\right\|_2^2,
\end{equation}
where $\phi_c^{\,l}$ and $\phi_r^{\,l}$ denote the feature representations of $I_c$ and $I_r$ at layer $l$, $(H_l,W_l)$ is the spatial size of that layer, and $w_l$ is the learned weight. Lower LPIPS values indicate higher perceptual similarity.

\subsection{Experimental settings}

The navigation-derived baseline was extracted using a Savitzky-Golay filter with a window length $2K+1$ of 301 and a polynomial order of 1. For image-based microscopic attitude inference, the stripe height and along-track stride were set to $h_s=6$ and $\Delta_y=2$, respectively. The multi-scale ROI setting used $\mathcal{S}=\{0.2,\,0.4,\,0.6,\,0.8\}$ and $\rho=0.5$. In the internal consistency score $E_{j,s,i}$, the weights were set to $\lambda=0.6$ and $\mu=0.3$. The strictness coefficient in adaptive stripe-wise distortion localization was set to $\kappa=0.75$; intervals shorter than 15 rows were removed and gaps shorter than 10 rows were merged. For port-starboard mode separation, the overlap threshold was set to $\tau_{ov}=0.5$. In the geocoding stage, empty grid cells were completed by normalized convolution using a $7\times7$ kernel $\mathcal{K}$.

\subsection{Visualization results}

\begin{figure*}[htbp]
	\centering
	\includegraphics[width=1\linewidth]{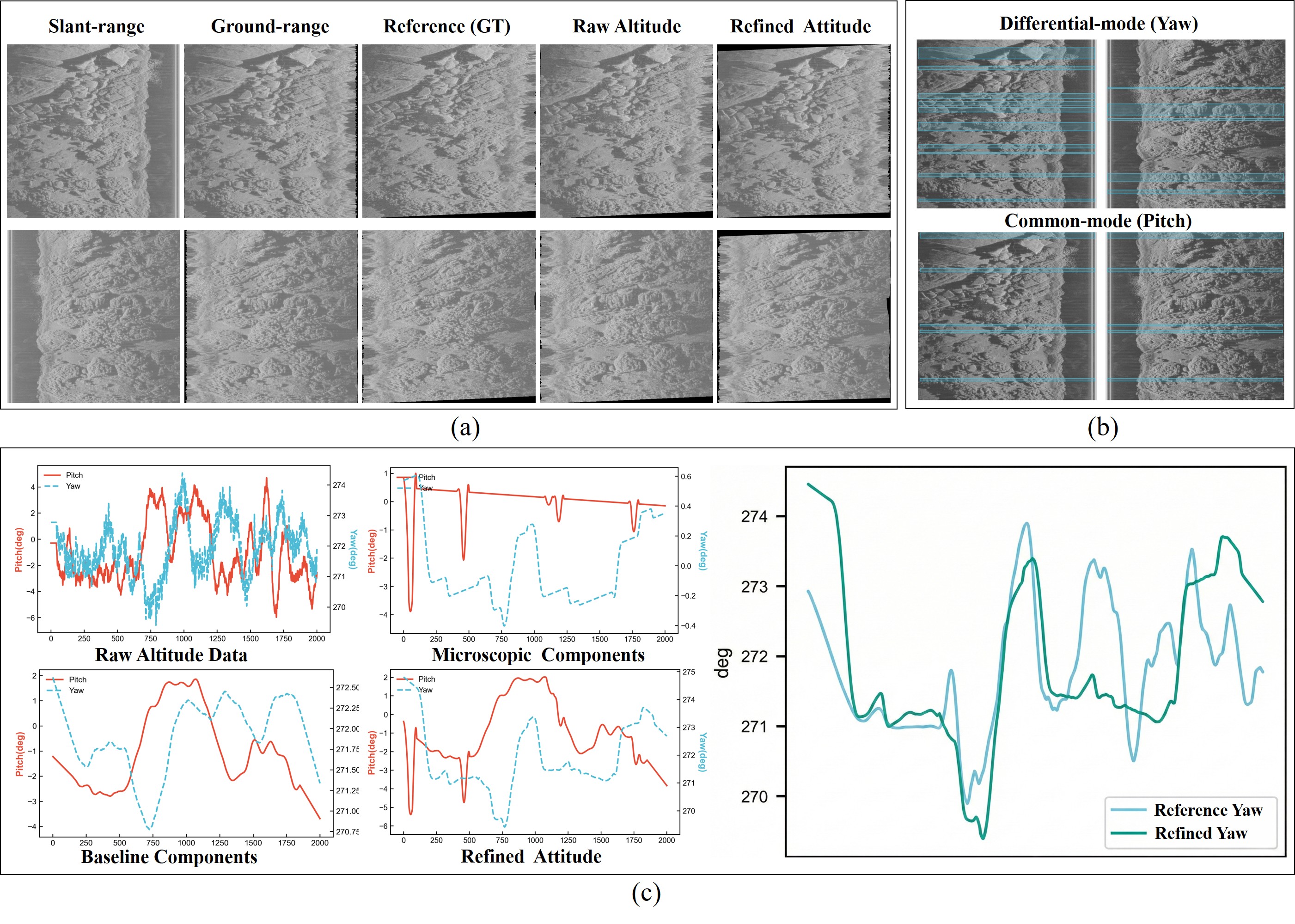}
	\caption{Visualization results on Dataset I. (a) Comparison of the slant-range image, ground-range image, reference image (GT), image geocoded with perturbed raw attitude, and image corrected with the refined attitude. The reference image is obtained by geocoding with the original reliable navigation and attitude data, while the raw-attitude result is obtained by adding noise and temporal delay to the original yaw and pitch sequences before geocoding. (b) Support separation results, where differential-mode supports correspond to yaw-related one-sided distortion responses and common-mode supports correspond to pitch-related synchronized responses. (c) Decomposition and refinement results of the attitude signals, including the raw attitude data, navigation-derived baseline, microscopic perturbations, refined attitude, and a comparison between the reference yaw and refined yaw.}
	\label{fig10}
\end{figure*}	
Fig. \ref{fig10} shows the visualization results on Dataset I. Although both Dataset I and Dataset II provide reliable navigation and attitude data, Dataset II is dominated mainly by yaw-induced distortion with only limited pitch influence; therefore, to simultaneously visualize and analyze the refinement effects on both yaw- and pitch-related distortions, we use Dataset I as the representative example. Since the navigation and attitude data in this dataset are reliable, the image geocoded with the original navigation and attitude parameters is taken as the reference image (GT). To simulate inaccurate navigation input for evaluation, the raw-attitude result was generated by perturbing the original yaw and pitch sequences with additional noise and temporal delay \cite{A29}, and then using the perturbed attitude in the same geocoding framework. As shown in Fig. \ref{fig10}(a), the ground-range result still contains evident geometric distortion, and the image geocoded with the perturbed raw attitude exhibits visible inter-line misalignment and local stretching. After replacing the raw attitude with the refined attitude sequence $\mathbf{\Theta}_{\mathrm{ref}}(n)$, these artifacts are visibly alleviated, and the corrected image becomes more consistent with the reference image in structural continuity.

As shown in \ref{fig10}(b), the differential-mode supports mainly appear as one-sided responses between the port and starboard swaths, consistent with the yaw-related distortion pattern, whereas the common-mode supports appear as synchronized responses on both sides, consistent with the pitch-related distortion pattern. Fig. \ref{fig10}(c) further shows the navigation-derived baseline, microscopic perturbations, and refined attitude sequence. The refined attitude preserves the low-frequency trend of the baseline component while introducing localized corrections in the detected distorted intervals. For the yaw example, the refined sequence captures several local variations from the reference data but does not achieve a full trajectory recovery.

\subsection{Robustness Study}

\begin{figure*}[htbp]
	\centering
	\includegraphics[width=1\linewidth]{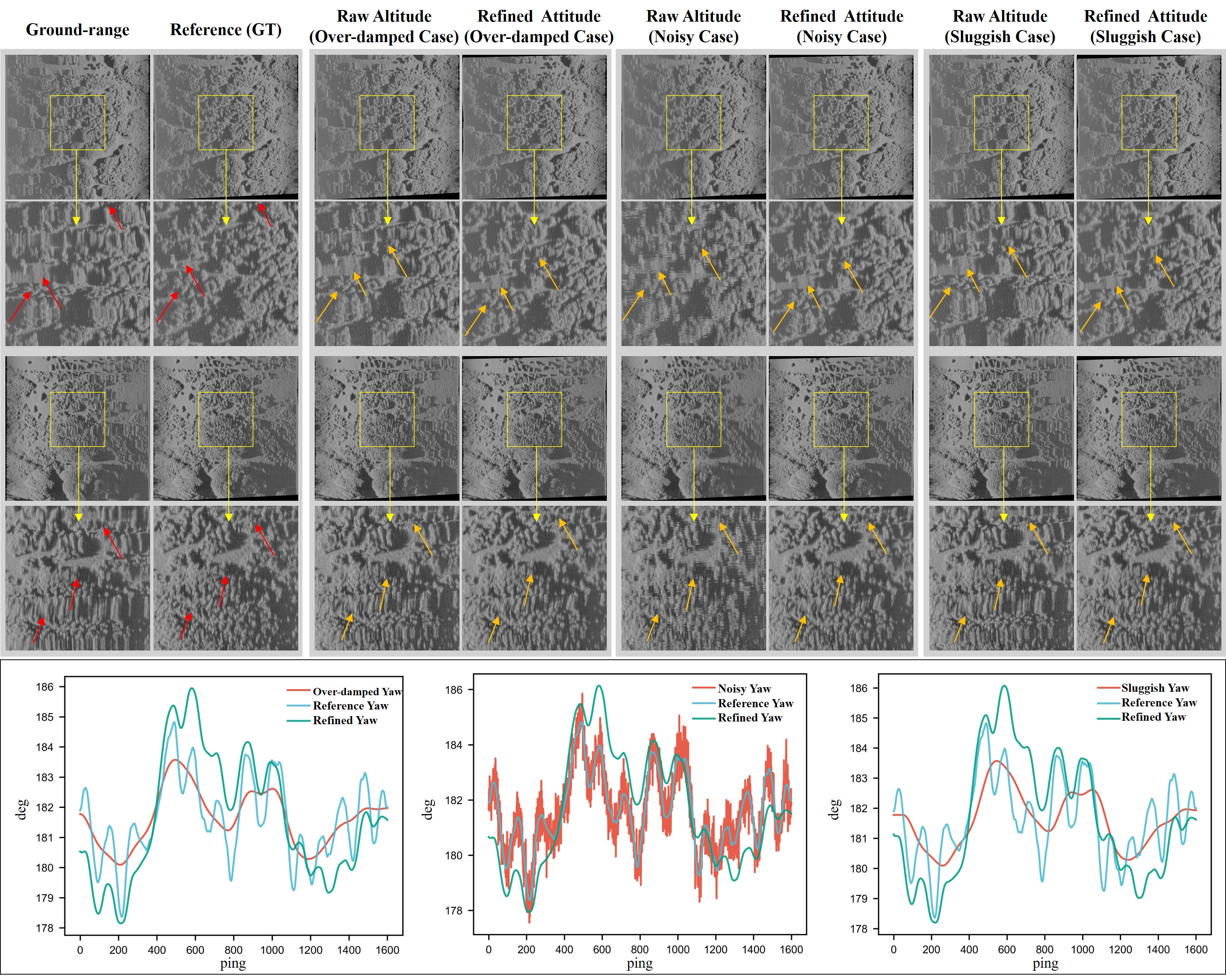}
	\caption{Robustness evaluation under degraded attitude input. The reference image (GT) is obtained by geocoding with the original reliable navigation and attitude data, while each raw attitude-based image is generated by using the corresponding degraded attitude sequence in the same geocoding framework. From left to right: \emph{Over-damped}, \emph{Noisy}, and \emph{Sluggish} cases, together with their corrected images based on refined attitude. Zoomed regions highlight local geometric differences. The bottom row compares the degraded yaw, reference yaw, and refined yaw for the three cases.}
	\label{fig11}
\end{figure*}	

\begin{figure*}[htbp]
	\centering
	\includegraphics[width=1\linewidth]{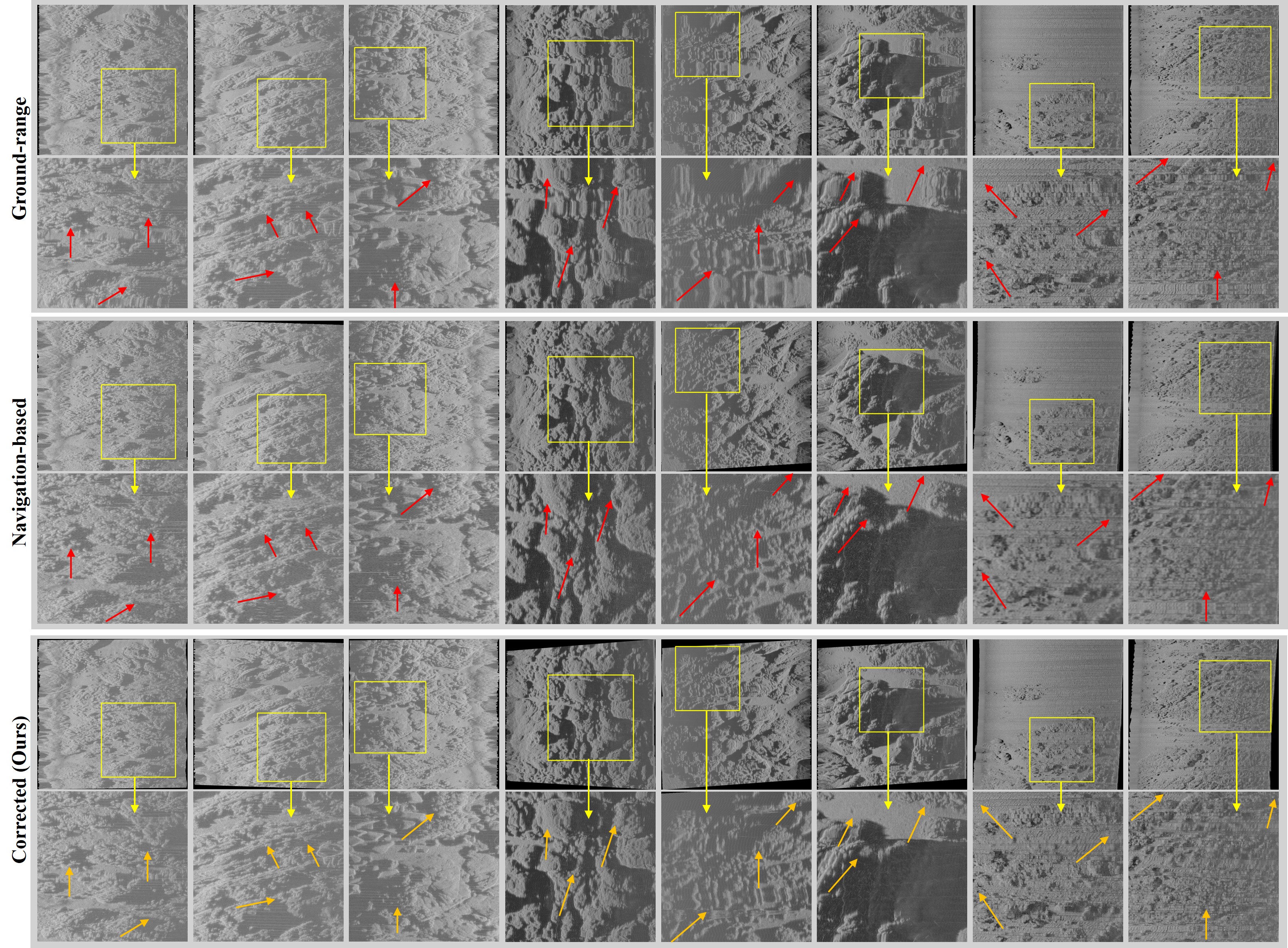}
	\caption{Generalization results on three datasets. Columns 1-3, 4-6, and 7-8 correspond to Dataset I, Dataset II, and Dataset III, respectively. The first row shows the distorted images in the ground-range representation for consistent visualization. The second row shows the navigation-based results obtained by geocoding with the recorded navigation and attitude data, and the third row shows the results corrected by the proposed method. Yellow boxes indicate enlarged regions, and arrows highlight typical local geometric differences. In the highlighted regions, the proposed method alleviates part of the local geometric artifacts, especially inter-line misalignment and structural discontinuity, and yields more coherent local appearance in most cases. Nevertheless, the improvement varies across samples, and some residual distortions remain, particularly in more challenging regions.}
	\label{fig12}
\end{figure*}	

\begin{table*}[t]
\centering
\begin{threeparttable}
\caption{Robustness evaluation under  degraded attitude input. We report image-quality metrics between corrected SSS images and GT.
Lower LPIPS is better ($\downarrow$); higher PSNR, MS-SSIM, and NCC are better ($\uparrow$). We additionally report the change
$\Delta = \text{Refined} - \text{Raw}$ for each metric.}
\label{robustness}
\renewcommand{\arraystretch}{1.2}
\setlength{\tabcolsep}{11.5pt} 

\small
\begin{tabular}{
  l l
  S[table-format=1.4]  S[table-format=+1.4]
  S[table-format=2.4]  S[table-format=+2.4]
  S[table-format=1.4]  S[table-format=+1.4]
  S[table-format=1.4]  S[table-format=+1.4]
}
\toprule
\multicolumn{2}{l}{\textbf{Setting}} &
\multicolumn{2}{c}{\makecell{\textbf{LPIPS}~(\(\downarrow\))}} &
\multicolumn{2}{c}{\makecell{\textbf{PSNR}~(\(\uparrow\))}} &
\multicolumn{2}{c}{\makecell{\textbf{MS-SSIM}~(\(\uparrow\))}} &
\multicolumn{2}{c}{\makecell{\textbf{NCC}~(\(\uparrow\))}} \\
\cmidrule(lr){3-4}\cmidrule(lr){5-6}\cmidrule(lr){7-8}\cmidrule(lr){9-10}
& & {\textbf{Value}} & {\(\boldsymbol{\Delta}\)} &
    {\textbf{Value}} & {\(\boldsymbol{\Delta}\)} &
    {\textbf{Value}} & {\(\boldsymbol{\Delta}\)} &
    {\textbf{Value}} & {\(\boldsymbol{\Delta}\)} \\
\midrule

\multirow{2}{*}{Over-damped}
 & Raw & 0.206 & {}      & 20.852 & {}      & 0.660 & {}      & 0.626 & {}      \\
 & Refined  & 0.196 & \textbf{-0.010} &  23.562 & \textbf{+2.710} & 0.772 & \textbf{+0.112} & 0.827 & \textbf{+0.201} \\
\addlinespace[3pt]
\midrule
\addlinespace[3pt]

\multirow{2}{*}{Noisy}
 & Raw & 0.268 & {}      & 21.624 & {}      & 0.770 & {}      & 0.819 & {}      \\
 & Refined  & 0.201 & \textbf{-0.067} & 23.241 & \textbf{+1.617}  & 0.765 & \textbf{-0.005} & 0.788 & \textbf{-0.031} \\
\addlinespace[3pt]
\midrule
\addlinespace[3pt]

\multirow{2}{*}{Sluggish}
 & Raw & 0.212 & {}      & 20.436 & {}      & 0.624 & {}      & 0.594 & {}      \\
 & Refined  & 0.204 & \textbf{-0.008} & 23.178 & \textbf{+2.742} & 0.748 & \textbf{+0.124} & 0.790 & \textbf{+0.196} \\
\bottomrule
\end{tabular}
\end{threeparttable}
\end{table*}

To evaluate the robustness of the proposed image-based microscopic attitude inference under degraded attitude input, we considered three representative cases, namely \emph{Over-damped} \cite{A30}, \emph{Noisy} \cite{A31}, and \emph{Sluggish} \cite{A32}. As in the visualization experiment, the reference image (GT) was obtained by geocoding with the original reliable navigation and attitude data, whereas each raw attitude-based image was generated by using the corresponding degraded yaw-pitch sequence in the same geocoding framework. The over-damped case suppresses high-frequency attitude variations, the noisy case adds random fluctuations, and the sluggish case combines attenuation with temporal delay.

Fig. \ref{fig11} shows both the corrected images and the corresponding yaw curves. For each case, the upper panels compare the ground-range image, the reference image, and the images geocoded from degraded and refined attitudes; the zoomed views highlight typical local distortions, where broken or oscillatory stripe patterns indicate geometric inconsistency. The lower plots show that the refined yaw recovers part of the missing local variation in the over-damped and sluggish cases, while suppressing part of the random fluctuation in the noisy case. Taking the over-damped case as an example, excessive smoothing removes the local yaw variation needed for consistent geocoding, leading to blurred and discontinuous stripe structures in the raw result. After refinement, part of this local variation is recovered, and the corresponding image shows improved structural continuity and reduced inter-line distortion. Similar improvement is observed in the sluggish case, whereas the noisy case remains more challenging because strong random perturbations are less consistently recoverable from image evidence alone.

The quantitative results in Table \ref{robustness} are consistent with these observations. The over-damped and sluggish cases show clear improvement after correction, with reduced LPIPS and increased PSNR, MS-SSIM, and NCC. In contrast, the noisy case is less stable, although LPIPS and PSNR improve after correction, MS-SSIM and NCC decrease slightly. This suggests that, under strong random perturbations, the proposed refinement can still alleviate part of the distortion and improve pixel-level fidelity, but its effect on structural similarity and global correlation is more limited.

\subsection{Generalization Study}

\begin{table}[t]
\centering
\begin{threeparttable}
\caption{Generalization performance on two datasets. We report image-quality metrics between (i) ground-range images and GT and (ii) corrected images and GT.}
\label{generalization}
\renewcommand{\arraystretch}{1.2} % Increase row height for readability
\setlength{\tabcolsep}{6.5pt}

\small
\begin{tabular}{
  l l
  S[table-format=1.3]
  S[table-format=2.3]
  S[table-format=1.3]
  S[table-format=1.3]
}
\toprule
\multicolumn{2}{l}{\textbf{Dataset}} 
& {\makecell{\textbf{LPIPS}\\(\(\downarrow\))}} 
& {\makecell{\textbf{PSNR}\\(\(\uparrow\))}} 
& {\makecell{\textbf{MS-SSIM}\\(\(\uparrow\))}} 
& {\makecell{\textbf{NCC}\\(\(\uparrow\))}} \\
\midrule

\multirow{2}{*}{Dataset 1}
  & Raw  & 0.227 & 20.034 & 0.617 & 0.587 \\
  & Refined  & \bfseries 0.211 & \bfseries 22.271 & \bfseries 0.751 & \bfseries 0.755 \\
\addlinespace[4pt]
\midrule
\addlinespace[4pt]

\multirow{2}{*}{Dataset 2}
   & Raw  & 0.250 & 18.039 & 0.561 & 0.521 \\
  & Refined  & \bfseries 0.230 & \bfseries 20.677 & \bfseries 0.702 & \bfseries 0.711 \\
\bottomrule
\end{tabular}
\end{threeparttable}
\end{table}

To further assess the generalization of the proposed method, we report results on the three datasets introduced above. In Fig. \ref{fig12}, the first three columns, the middle three columns, and the last two columns correspond to Dataset I, Dataset II, and Dataset III, respectively. The ground-range image denotes the distorted data projected onto the ground plane, which is used solely for consistent visual comparison without the water-column region. The navigation-based result is obtained by geocoding with the recorded attitude data, while the last row shows the result corrected by the proposed method.

shown in Fig. \ref{fig12}, the yellow boxes and enlarged views highlight representative local geometric artifacts, mainly including stripe discontinuity, inter-line staggering, and oscillatory deformation. Taking the first column as an example, the ground-range image shows evident local waviness and structural breakup. After geocoding with the recorded navigation data, the overall structure becomes substantially more organized, indicating that the main geometric distortion has already been effectively corrected. The result produced by the proposed method remains visually close to this navigation-based image, while showing slightly more regular local stripe arrangement in some highlighted details. Similar tendencies can be observed in most samples across the three datasets, although the degree of improvement varies and some residual artifacts still remain. For Dataset I and Dataset II, quantitative comparisons against the reference images (navigation-based results) are reported in Table \ref{generalization}. In both datasets, the corrected images achieve lower LPIPS and higher PSNR, MS-SSIM, and NCC than the distorted ground-range images. For Dataset III, only qualitative comparison is provided. In this case, the navigation-based result still contains noticeable local geometric artifacts, and the proposed method can alleviate part of them, although the improvement is not uniform for all samples.

\section{Conclusion}

This paper introduced an image-consistent geometric correction framework for SSS imagery. The results show that incorporating image-derived microscopic refinement into navigation-based correction can effectively improve local geometric consistency while preserving global geometric referencing. Experiments on real SSS datasets further demonstrate that the proposed method reduces typical geometric artifacts, including inter-line misalignment, local stretching, and structural discontinuity, and remains effective under both degraded-attitude and cross-dataset evaluation settings. Two limitations should be noted. First, the current framework still depends on several empirically selected parameters and thresholds, which may reduce robustness when imaging conditions or seabed scenes vary significantly. Second, the present refinement mainly targets stretching-dominant distortion patterns, while compression-like changes are not explicitly modeled and remain difficult to distinguish from terrain-induced structural variation based on image appearance alone. Future work will focus on improving parameter adaptability and extending the framework to handle a broader range of local distortion patterns in more challenging seabed environments.

\section*{Acknowledgments}

This work was supported in part by the National Natural Science Foundation of China (Grant No. 62171368) and the Science, Technology and Innovation Commission of Shenzhen Municipality (Grant Nos. JCYJ20241202124931042, ZDCYKCX20250901093900002), and in part by the Spanish government through projects ASSiST (PID2023-149413OB-I00) and IURBI (CNS2023-144688).

Can Lei also gratefully acknowledges the scholarship support from the China Scholarship Council (CSC).

\bibliographystyle{IEEEtran}
\bibliography{refernew}

\vfill

\end{document}